\documentclass[12pt]{article}

\newtheorem{pro}{Proposition}

\usepackage{times,epsfig,graphics,amssymb,latexsym,amsmath,setspace,fullpage}
\usepackage{array,threeparttable,hyperref,url,lscape}
\usepackage[usenames]{color}

\def\boxit#1{\vbox{\hrule\hbox{\vrule\kern6pt\vbox{\kern6pt#1\kern6pt}\kern6pt\vrule}\hrule}}

\def\argmin{\mathop{\rm argmin}}

\def\sign{\mathop{\rm sign}}

\begin{document}
\pagenumbering{arabic}

\setcounter{page}{1}

\title{\Large \bf Flexible combination of multiple diagnostic biomarkers to improve diagnostic accuracy}

\author{ Tu Xu$^\dag$, Yixin Fang$^\ddagger$, Alan Rong$^\sharp$,  Junhui Wang$^\natural$\\[10pt]
  $^\dag$ Gilead Sciences Inc.\\
     \and
    $^\ddag$ Division of Biostatistics\\
        Department of Population Health\\
        New York University\\
     \and
~~~~~~~~~~~~~~~~~~~~~$^\sharp$ Astellas Pharma Inc.~~~~~~~~~~~~~~~~~~~\\[2pt]
          \and
   ~~~~~~~$^\natural$ Department of Mathematics~~~~~~~~ \\
          City University of Hong Kong }

\date{}
\maketitle

\begin{abstract}
In medical research, it is common to collect information of multiple continuous biomarkers to improve the accuracy of diagnostic tests. Combining the measurements of these biomarkers into one single score is a popular practice to integrate the collected information, where the accuracy of the resultant diagnostic test is usually improved. To measure the accuracy of a diagnostic test, the Youden index has been widely used in literature.  Various parametric and nonparametric methods have been proposed to linearly combine biomarkers so that the corresponding Youden index can be optimized. Yet there seems to be little justification of enforcing such a linear combination. This paper proposes a flexible approach that allows both linear and nonlinear combinations of biomarkers. The proposed approach formulates the problem in a large margin classification framework, where the combination function is embedded in a flexible reproducing kernel Hilbert space. Advantages of the proposed approach are demonstrated in a variety of simulated experiments as well as a real application to a liver disorder study.

\end{abstract}

\vskip .1 in
\noindent {\small {\bf Key words}: \it biomarker, diagnostic accuracy, margin, receiver operating characteristic curve, reproducing kernel Hilbert space, Youden index}

\doublespacing

\section{Introduction}

In medical research, continuous biomarkers have been commonly explored as diagnostic tools to distinguish subjects, such as diseased and non-diseased groups \cite{Shapiro1999}. The accuracy of a diagnostic test is usually evaluated through sensitivity and specificity, or the probabilities of true positive and true negative for any given cut-point. Particularly, the receiver operating characteristic (ROC) curve is defined as sensitivity versus $1-$specificity over all possible cut-points for a given biomarker \cite{Zhou2002,Pepe2003}, which is a comprehensive plot that displays the influence of a biomarker as the cut-point varies. To summarize the overall information of an ROC curve, different summarizing indices have been proposed, including the Youden index \cite{Youden1950} and the area under the ROC curve (AUC; \cite{Bamber1975}).

The Youden index, defined as the maximum vertical distance between the ROC curve and the $45^{\circ}$ line, is an indicator of how far the ROC curve is from the uninformative test \cite{Pepe2003}. Normally, it ranges from $0$ to $1$ with $0$ for an uninformative test and $1$ for an ideal test. The Youden index has been successfully applied in many clinical studies and served as an appropriate summary for the diagnostic accuracy of a single quantitative measurement (e.g., \cite{Zhou2002,Aoki1997,Perkins2006}).

It has been widely accepted by medical researchers that diagnosis based on one single biomarker may not provide sufficient accuracy \cite{Sidransky2002, Kumar2006}. Consequently, it is becoming more and more common that multiple biomarker tests are performed on each individual, and the corresponding measurements are combined into one single score to help clinicians make better diagnostic judgment. In literature,  various statistical modeling strategies have been proposed to combine biomarkers in a linear fashion. For instance, Su and Liu \cite{Su1993} derived the analytical results of optimal linear combination based on AUC under multivariate normal assumption. Pepe and Thompson \cite{Pepe2000} proposed to relax the distributional assumption and perform a grid search for the optimal linear combination, while its computation becomes expensive when the number of biomarkers gets large. Recently, a number of alternatives were proposed to alleviate the computational burden. For instances, the min-max approach \cite{Liu2011} combines only the minimum and maximum values of biomarker measurements linearly; the stepwise approach \cite{Kang2013} combines all biomarker measurements in a stepwise manner. By targeting directly on the optimal diagnostic accuracy, Yin and Tian \cite{Tian2014} extended these two methods to optimize the Youden index and demonstrated their improved performance in a number of numerical examples.

In recent years, nonlinear methods have been popularly employed to combine multiple biomarkers in various fields, including genotype classification \cite{Kouskoumvekaki2008}, medical diagnosis \cite{Chris2009}, and treatment selection \cite{Huang2014}.  In this paper, a new model-free approach is proposed and formulated in a large margin classification framework, where the biomarkers are flexibly combined into one single diagnostic score so that the corresponding Youdex index \cite{Youden1950} is maximized. Specifically, the combination function is modeled non-parametrically in a flexible reproducing kernel Hilbert space (RKHS; \cite{Wahba1990}), where both linear and nonlinear combinations could be accommodated via a pre-specified kernel function.

The rest of the paper is organized as follows. In Section 2, we provides some preliminary background of combining multiple biomarkers based on the Youden index. In Section 3, we discuss the motivation for flexible combinations and formulate the proposed flexible approach in a framework of large margin classification for combining multiple biomarkers. In Section 4, we conduct numerical experiments to demonstrate the advantages of the proposed approach. In Section 5, we apply the proposed approach to a liver disorder study. Section 6 contains some discussion.

\section{Preliminaries}
Suppose that every subject has $m$ biomarker measurements ${\bf X} = (X_{(1)}, X_{(2)}, \ldots, X_{(m)})$ with a probability density function $f({\bf X})$, where $X_{(j)}$ is a continuous measurement of the $j$-th biomarker. It also has a binary response variable $Y \in \{1,-1\}$ indicating the subject is diseased or not. In literature, researchers from different fields \cite{Sidransky2002, Kumar2006, Tian2014} have discussed and explored the validity of combining $m$ biomarker measurements into one single score function $g({\bf X})$ as a more powerful diagnostic tool. A subject is diagnosed as diseased if the combined score $g({\bf X})$ is higher than a given cut-point $c$, and non-diseased otherwise. To summarize its diagnostic accuracy, the Youden index is commonly used in practice. With sensitivity and specificity defined as $\mbox{sen}(g,c)=Pr(g({\bf X}) \geq c|Y=1)$ and $\mbox{spe}(g,c)=Pr(g({\bf X}) < c|Y=-1)$ respectively, the Youden index is formulated as
$$
J= \max_{g,c}~\{\mbox{sen}(g,c) + \mbox{spe}(g,c)- 1\}.
$$
The Youden index normally ranges from 0 to 1, where $J=1$ corresponds to a perfect separation, and $J=0$ corresponds to a random guess. 

To estimate the Youden index, various modeling strategies have been proposed. Schisterman et al.~\cite{Schisterman2005} provided a closed form for the Youden index assuming the conditional distribution of ${\bf X}|Y=\pm 1$ follows a multivariate Gaussian distribution. Further relaxing the distributional assumption, kernel smoothing techniques were adopted by Yin and Tian  \cite{Tian2014} and Fluss et al. \cite{Fluss2005}, where the sensitivity and specificity were estimated in a nonparametric fashion. 

Note that the formulation of $J$ can be rewritten as 
\begin{eqnarray} 
J  &=& \max_{g,c}~w(1)Pr \big(g({\bf X}) \geq c, Y=1 \big) + w(-1) Pr \big(g({\bf X}) < c, Y = -1 \big) -1 \nonumber \\
&=& \max_{g,c}~ \frac{1}{2} E\Big(w(Y) \big(1 + Y \sign(g({\bf X})-c)\big)\Big)-1, \label{eqn:YoudenJ}
\end{eqnarray}
where $w(1)=1/\pi$, $w(-1)=1/(1-\pi)$, $\pi=Pr(Y=1)$, and $\sign(u) = 1$ if $u \geq 0$ and $-1$ otherwise.  Denote the ideal combination function $g^*({\bf x})$ and cut-point $c^*$ as the ones that maximize $J$ over all possible functionals and cut-points. Following the proof of Proposition 1 in \cite{Xu2014}, the ideal $g^*({\bf x})$ and $c^*$ must satisfy 
\begin{equation}
\sign(g^*({\bf x}) - c^*) = \sign\left(p({\bf x}) - \pi\right),
\label{eqn:bayes}
\end{equation}
where $p({\bf x}) = Pr(Y=1|{\bf x})$ is the conditional probability of disease given the biomarker measurements.

 \section{Linear or nonlinear combination}

In (\ref{eqn:bayes}), the ideal $g^*({\bf x})$ and $c^*$ are defined based on $p({\bf x})$ that
is often unavailable in practice. Hence the expectation in (\ref{eqn:YoudenJ}) needs to be estimated based on the given sample $({\bf x}_i,y_i)_{i=1}^n$. Specifically, a natural estimate $\hat J$ can be obtained as
\begin{eqnarray}
	\hat J &=& \max_{g,c}~ \frac{1}{n} \sum_{i=1}^n \hat w(y_i)(1 + y_i \sign(g({\bf x}_i)-c)) -1 \nonumber \\
	&=& \max_{g,c}~ \frac{1}{|{\cal S}_1|}\sum\limits_{i \in {\cal S}_1} (1+\sign(g({\bf x}_i)-c)) + \frac{1}{|{\cal S}_{-1}|}\sum\limits_{i \in {\cal S}_{-1}} (1-\sign(g({\bf x}_i)-c))-1,
	\label{eqn:YoudenJ_est}
\end{eqnarray}
where $\hat w(1)=1/\hat \pi=n/|{\cal S}_1|$, $\hat w(-1)=n/|{\cal S}_{-1}|$, ${\cal S}_1 = \{i:y_i = 1\}$, ${\cal S}_{-1} = \{i: y_i = -1\}$, and $|\cdot|$ denotes the set cardinality. 

The optimization in (\ref{eqn:YoudenJ_est}) is generally intractable without a specified candidate space of $g$. In literature, linear functional space $g({\bf x}) = {\bf \beta}^T {\bf x}$ is often used \cite{Su1993,Pepe2000,Liu2011,Kang2013,Tian2014}, mainly due to its convenient implementation and natural interpretation. Yet there seems to be lack of scientific support for the use of linear combination of biomarkers. 

%\subsection{Rational for Nonlinear Combination}

%According to (\ref{eqn:bayes}), the ideal $g^*$ and $c^*$ must satisfy that $\sign(g^*({\bf X}-c^*)) = \sign(p({\bf X}) - \pi)$, and therefore the linear combination would be optimal if  $p({\bf x}) \geq \pi$ can be expressed in a linear form of ${\bf x}$. 

Consider a toy example, where $\pi = 1/2$, ${\bf X}|Y=1 \sim N_2((1,1)^T,I_2)$ and ${\bf X}|Y=-1 \sim N_2((0,0)^T,I_2)$, where $I_2$ is a 2-dimensional identity matrix. Then for any given ${\bf x}$,
$$
p({\bf x}) = \frac{f({\bf x}|Y=1)}{f({\bf x}|Y=1)+f({\bf x}|Y=-1)} = \frac{1}{1+e^{1-(x_{(1)}+x_{(2)})}}.
$$ 
where ${\bf x}=(x_{(1)},x_{(2)})^T$. Thus, the ideal combination of biomarkers $g^*({\bf x})$ can take the linear form $g^*({\bf x})=x_1+x_2$, leading to $\sign(g^*({\bf x}) - c) = \sign\big( p({\bf x}) - 1/2 \big )$ with $c=1$.  However, if the biomarkers are heterocedastic in the positive and negative groups, the ideal combination would be no longer linear. 
For instance, when ${\bf X}|Y=1 \sim N_2((1,1)^T,I_2)$ but ${\bf X}|Y=-1 \sim N_2((0,0)^T,2I_2)$, 
$$
p({\bf x}) = \frac{f({\bf x}|Y=1)}{f({\bf x}|Y=1)+f({\bf x}|Y=-1)} = \frac{2}{2+e^{1-(x_{(1)}+x_{(2)})+(x_{(1)}^2+x_{(2)}^2)/4}} .
$$ 
Clearly, the ideal combination of biomarkers is a quadratic function $g^*({\bf x})= \frac{x_{(1)}^2+x_{(2)}^2}{4}-(x_{(1)}+x_{(2)})$ with $c=\log(2) - 1$. Furthermore, if the conditional distribution ${\bf X}|Y$ is unknown, then the ideal combination of biomarkers may take various forms, and thus a pre-specified assumption on linear combination can be too restrictive and lead to suboptimal combinations.

%\section{A new estimation framework}

%The optimal solution to (\ref{eqn:YoudenJ_est}) is searched within a reproducing kernel Hilbert space ${\cal H}_K$ (RKHS; \cite{Wahba1990}), which includes both linear and non-linear functionals of $g(x)$. 

\subsection{Model-free estimation formulation}

\begin{figure}[!h]
	\caption{The 0-1 loss function, $\psi$ loss and $\psi_{0.5}$ loss.}
	\begin{center}
		\includegraphics[width=0.75\textwidth]{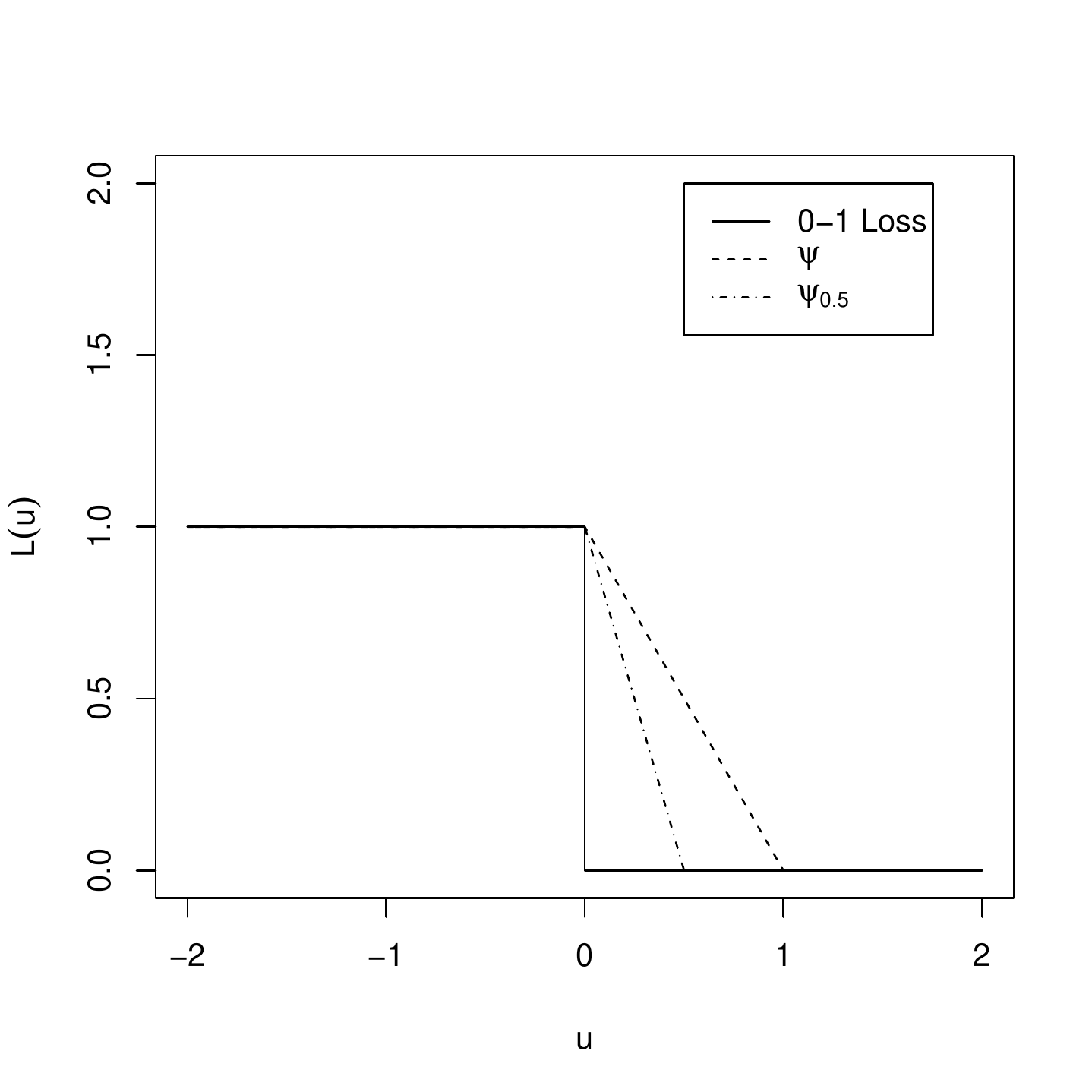}
		\label{fig:loss_function}
	\end{center}
\end{figure}

To allow more flexible $g({\bf x})$ than linear functions, it is natural to optimize (\ref{eqn:YoudenJ_est}) over a bigger functional space consisting of nonlinear functions. Note that the objective function in (\ref{eqn:YoudenJ_est}) involves a $\sign$ operator, which makes it discontinuous in $g$ and thus difficult to optimize in general \cite{Shen2003}. Alternatively, note that (\ref{eqn:YoudenJ_est}) can be simplified as
$$
\min_{g,c} ~ \frac{1}{n} \sum_{i=1}^n \hat w(y_i) \big (1 - \sign(u_i) \big ),
$$ 
where $u_i=y_i(g({\bf x}_i)-c)$. As proposed in Xu et al. \cite{Xu2014}, a surrogate $\psi_{\delta}$-loss, defined as 
$$
L_{\delta}(u)=\min \left\lbrace \frac{1}{\delta}(\delta-u)_+,1 \right \rbrace,
%=\frac{1}{\delta}(\delta-u)_+ - \frac{1}{\delta}(-u)_+., 
$$
can be employed to replace the 0-1 loss $L_{01}(u)=1-\sign(u)$ in the objective function. The $\psi_{\delta}$-loss extends the $\psi$-loss \cite{Shen2003,Liu2006} by introducing a parameter $\delta$ that controls the difference between the surrogate loss and the 0-1 loss. Figure \ref{fig:loss_function} displays the 0-1 loss, the $\psi$-loss and the $\psi_{0.5}$-loss as functions of $u$. 

\begin{center}
\begin{tabular}{c}
\hline
\hline
Figure \ref{fig:loss_function} about here.\\
\hline
\hline
\end{tabular}
\end{center}

Furthermore, denote ${\cal D}_{g,c,\epsilon} = \{{\bf x}: g({\bf x})-c \geq 0 \ \mbox{and} \  |p({\bf x}) -\pi| \geq \epsilon\}$. Proposition \ref{lem:consistency} shows that for any $\epsilon >0$, the $\psi_{\delta}$-loss is asymptotically Fisher consistent in estimating ${\cal D}_{g^*,c^*,\epsilon}$ when $\delta$ approaches 0. 

{\pro
\label{lem:consistency}
Given any $\epsilon>0$, let $(g^*_{\delta}, c^*_{\delta})=\argmin_{g,c} E \big ( w(Y) L_{\delta}( Y(g({\bf X})-c) ) \big )$, then as $\delta \rightarrow 0$,
$$
Pr\left( {\cal D}_{g^*_{\delta}, c^*_{\delta},\epsilon} \triangle {\cal D}_{g^*,c^*,\epsilon}\right) \rightarrow 0,
$$ 
where $\triangle$ denotes the symmetric difference of two sets.}

%needs to be optimized with respect to functional $g(x)$. It can no longer be solved by the exhaustive grid search as it is impossible to enumerate the uncountable functionals. In addition, (\ref{eqn:YoudenJ_est}) involves the 0-1 loss $L_{01} (u)=\frac{1}{2}(1-\sign(u))$, which is known to be intractable by any efficient optimization techniques \cite{Shen2003}. 

With the $\psi_{\delta}$-loss, the proposed model-free estimation framework for $(g({\bf x}),c)$ is formulated as
\begin{equation}
\min_{g \in {\cal H}_K,c \in \mathbb{R}} \\ \frac{1}{n} \sum_{i=1}^n \hat{w}(y_i) L_{\delta}( y_i(g({\bf x}_i)-c) ) + \lambda {\cal J}(g),
\label{eqn:rkhs}
\end{equation}
where $\lambda$ is a tuning parameter, ${\cal H}_K$ is set as a RKHS associated with a pre-specified kernel function $K(\cdot,\cdot)$, and ${\cal J}(g)=\frac{1}{2} \|g\|^2_{{\cal H}_K}$ is the RKHS norm penalizing the complexity of $g({\bf x})$. The popular kernel functions include the linear kernel $K({\bf u,v)=u}^T {\bf v}$, the $m$-th order polynomial kernel $K({\bf u,v})=(1+{\bf u}^T {\bf v})^m$, and the Gaussian kernel $K({\bf u,v})=\exp \{-\|{\bf u-v}\|^2/2\tau^2 \}$ with a scale parameter $\tau^2$. When the linear kernel is used, the resultant ${\cal H}_K$ contains all linear functions; when the Gaussian kernel is used, ${\cal H}_K$ becomes much richer and admits more flexible nonlinear functions.

More interestingly, the representer theorem \cite{Wahba1990} implies that the solution to (\ref{eqn:rkhs}) must be of the form $\hat g({\bf x}) = \sum_{i=1}^n a_i K({\bf x}_i,{\bf x})$, and thus $\| g\|_{{\cal H}_K}^2={\bf a}^T {\bf K} {\bf a}$ with ${\bf a}=(a_1,\cdots,a_n)^T$ and ${\bf K}=(K({\bf x}_i,{\bf x}_j))_{i,j=1}^n$. The representor theorem greatly simplifies the optimization task by turning the minimization over a functional space into the minimization over a finite-dimensional vector space. Specifically, the minimization task in (\ref{eqn:rkhs}) becomes
\begin{equation}
\min_{a \in \mathbb{R}^n,c \in \mathbb{R}}~ s(\tilde a) = \frac{1}{n} \sum_{i=1}^n \hat{w}_i(y_i) L_{\delta} \Big ( y_i \Big (\sum_{j=1}^n a_j K({\bf x}_i,{\bf x}_j)-c \Big ) \Big ) + \frac{\lambda}{2} {\bf a}^T {\bf K a},
\label{eqn:represent}
\end{equation}
where ${\bf \tilde{a}}=({\bf a}^T,c)^T$ is an $(n+1)$-dim vector. 

The minimization task in (\ref{eqn:represent}) involves a non-convex function $L_{\delta}(\cdot)$, and thus we employ the difference convex algorithm (DCA; \cite{An1997})  to tackle the non-convex optimization task. The DCA decomposes the non-convex objective function in to the difference of two convex functions, and iteratively approximates it through a refined convex objective function. It has been widely used for non-convex optimization and delivers superior numerical performance \cite{Liu2005,Xu2014,Huang2014}. The detail of solving \eqref{eqn:represent} is similar to that in \cite{Xu2014} and thus omitted here.

%The final estimation formulation then becomes
%\begin{equation}
%\min_{\substack{g \in {\cal H}_K\\c \in \mathbb{R}}}~\frac{1}{n} \sum_{i=1}^n \hat{w}(y_i) L_{\delta}( y_i(g(x_i)-c) )  + \frac{\lambda}{2} \|g\|^2_{{\cal H}_K},
%\label{eqn:rkhs}
%\end{equation}
%where ${\cal H}_K$ is the RKHS induced by some pre-specified kernel function $K(\cdot,\cdot)$ such as linear kernel or Gaussian kernel, and ${\cal J}(g)=\frac{1}{2}\| g\|_{{\cal H}_K}^2$ is the associated RKHS norm of $g(x)$. 

\section{Simulation examples}

This section examines the proposed estimation method for combining biomarkers in a number of simulated examples. The numerical performance of the proposed kernel machine estimation (KME) method is compared against some existing popular alternatives, including the min-max method (MMM) \cite{Liu2011}, the parametric method under multivariate normality assumption (MVN) \cite{Schisterman2007}, the non-parametric kernel smoothing method (KSM) with Gaussian kernel \cite{Tian2014}, the stepwise method (SWM) \cite{Kang2013}, and the other two classification methods in \cite{Kouskoumvekaki2008}, the logistic regression (LR) and the classification tree (TREE).

For illustration, the kernel function used in all methods is set as the linear kernel $K({\bf z}_1,{\bf z}_2)={\bf z}_1^{T} {\bf z}_2$ and the Gaussian kernel $K({\bf z}_1,{\bf z}_2)=e^{-\|{\bf z}_1-{\bf z}_2\|^2/2 \tau^2}$, where the scale parameter $\tau^2$ is set as the median of pairwise Euclidean distances between the positive and negative instances within the training set. The tuning parameter $\lambda$ for our proposed method is selected by 5-fold cross validation that maximizes the empirical Youden index
\begin{equation}
\tilde{J} = \frac{1}{5}\sum\limits_{k = 1}^{5} \left(\frac{\sum\limits_{i \in V_k } I(y_i=-1)I(\hat{g}({\bf x}_i) \leq c)}{\sum\limits_{i \in V_k} I(y_i=-1)} - \frac{\sum\limits_{i \in V_k} I(y_i=1) I(\hat{g}({\bf x}_i) \leq c)}{\sum\limits_{i \in V_k} I(y_i=1)}\right),
\label{eqn:empirical_J}
\end{equation}
where $I(\cdot)$ is an indicator function and $V_k$ is the validation set of $k$-th folder. The maximization is conducted via a grid search, where the grid for selecting $\lambda$ is set as $\{10^{(s-41)/10}; s=1,\cdots,81\}$. The optimal solutions of MVN and KSM are searched by routine {\it optim()} in R as suggested in Ying and Tian \cite{Tian2014}. SWM and MMM are based on the grid search with the same grid. TREE is tuned by default in R. Furthermore, for the proposed KME method, $\delta$ is set as $0.1$ for all simulated examples as suggested in Hedayat et al. \cite{Hedayat2014}.

Four simulated examples are examined. Example 1 is similar to Example 5.1.1 in \cite{Tian2014}. Example 2 modifies Example 1 by using multivariate Gamma distribution, which appears to be a popular model assumption in literature \cite{Schisterman2005}.
Examples 3 and 4 are similar to Setting 2 in \cite{Huang2014} and Example II(b) in \cite{Fong2014}, which simulate data from logistic models with nonlinear effect terms. 
  
{\it Example 1.} A random sample $\{({\bf X}_i,Y_i); i=1,\cdots, n\}$ is generated as follows. First, $Y_i$ is generated from $\mbox{Bernoulli}(0.5)$. Second, if $Y_i = 1$, then ${\bf X}_i$ is generated from $\mbox{MVN} \big({\boldsymbol \mu}_1, \Sigma_1 \big)$, where ${\boldsymbol \mu}_1 = (0.4, 1.0, 1.5, 1.2)^T$ and $\Sigma_1 = 0.3 I_4 + 0.7 J_4$ with $I_4$ a $4$-dimensional identity matrix and $J_4$ a $4 \times 4$ matrix of all $1$'s; if $Y_i= -1$, then ${\bf X}_i$ is generated from $\mbox{MVN} \big({\boldsymbol \mu}_2, \Sigma_1 \big)$ with ${\boldsymbol \mu}_2 = (0, 0, 0, 0)^T$.

{\it Example 2.} A random sample $\{({\bf X}_i,Y_i); i=1,\cdots, n\}$ is generated as follows. First, $Y_i$ is generated from $\mbox{Bernoulli}(0.5)$. Second, if $Y_i = 1$, then ${\bf X}_i$ is generated from a multivariate gamma distribution with mean ${\boldsymbol \mu}_1 = (0.55,0.7,0.85,1)^T$ and covariance matrix 
$\Sigma_1 = 0.25 J_4+\mbox{diag}(0.025,0.1,0.175,0.25)$;
%\bigl(\begin{smallmatrix} 0.275&0.250&0.250&0.250\\ 0.250&0.350 &0.250&0.250 \\ 0.250&0.250&0.425&0.250\\0.250&0.250&0.250&0.500 \end{smallmatrix} \bigr)$;
if $Y_i = -1$, then ${\bf X}_i$ is generated from multivariate gamma distribution with mean ${\boldsymbol \mu}_2 = (0.55,0.55,0.55,0.55)^T$ and covariance matrix 
$\Sigma_2 = 0.025 I_4 +  0.25 J_4$. The multivariate gamma distributed samples are generated with normal copula. 
%\bigl(\begin{smallmatrix} 0.275&0.250&0.250&0.250\\ 0.250&0.275 &0.250&0.250 \\ 0.250&0.250&0.275&0.250\\0.250&0.250&0.250&0.275 \end{smallmatrix} \bigr)$.

{\it Example 3.}  A random sample $\{({\bf X}_i,Y_i); i=1,\cdots, n\}$ is generated as follows. First, ${\bf X}_i$ is generated from $\mbox{MVN} \big({\boldsymbol \mu}, \Sigma \big)$, where ${\boldsymbol \mu} = (0,0,0,0)^T$ and $\Sigma = 0.3I_4 + 0.7 J_4$. Second, $Y_i$ is generated from a logistic model with $\mbox{logit}(p({\bf x})) = x_{(1)} + x_{(2)}^2 + x_{(3)}^3 + x_{(4)}^4 - 1.5$.

{\it Example 4.}  A random sample $\{({\bf X}_i,Y_i); i=1,\cdots, n\}$ is generated as follows. First, ${\bf X}_i$ is generated from $t_4 \big({\boldsymbol \mu}, \Sigma \big)$, where ${\boldsymbol \mu} = (0,0,0,0)^T$ and $\Sigma = I_4$. Second, $Y_i$ is generated from a logistic model with $\mbox{logit}(p({\bf x})) = 8\big( \mbox{sin}(0.5\pi x_{(1)}) + \mbox{cos}(\pi x_{(1)} x_{(2)})+ x_{(3)}^2$ + $3x_{(3)} x_{(4)} + x_{(4)}^2\big).$

In all examples, the sample sizes for training $n_{tr}$ and testing $n_{te}$ are set as $n_{tr}=100, 250, 500$ and $n_{te} = 2000$, respectively. Each scenario is replicated 100 times. The averaged empirical Youden index $\hat{J}$, as well as the corresponding standard deviations, are summarized in Table \ref{tab:optimal_J}.

\begin{center}
\begin{tabular}{c}
\hline
\hline
Table \ref{tab:optimal_J} about here.\\
\hline
\hline
\end{tabular}
\end{center}

It is evident that our proposed methods, linear kernel machine estimation method  (LKME) and Gaussian kernel machine estimation method (GKME), yield competitive performance in all examples. The performance of MVN, SWM, and LR is competitive in Example 1 as the data within each class indeed follows a Gaussian distribution sharing a common covariance structure, and thus the linear combination is optimal. Their performance becomes less competitive in other examples when linear combination is no longer optimal. It is evident that in Examples 3 and 4, with nonlinear patterns specified, the GKME outperforms all other methods. Especially, in Example 4, the performance of GKME is outstanding due to a strong nonlinear pattern specified. In general, the performance of KSM is less competitive. It could be due to the over-fitting issue when applying the Gaussian kernel to estimate sensitivity and specificity. With similar exhaustive grid search, the performance of SWM is better than MMM in Examples 1 and 4 but worse in Examples 2 and 3. As for the two classification methods, LR yields competitive performance in Examples 1 and 2 and becomes less competitive when logistic models with nonlinear patterns are applied in Examples 3 and 4. The performance of TREE is modest considering the nature of recursive partition. 

\section{Real application}

In this section, our proposed method is applied to a study of liver disorder. The dataset consists of 345 male subjects with 200 subjects in the control group and 145 subjects in the case group. For each subject, there are five blood tests (mean corpuscular volume, alkaline phosphotase, alamine aminotransferase, aspartate aminotransferase, and gamma-glutamyl transpeptidase) which are thought to be sensitive to liver disorders that may be related to excessive alcohol consumption, and another covariate with the average daily alcoholic beverages consumption information. The corresponding empirical estimates of the Youden index of all six markers are 0.141, 0.178, 0.174, 0.144, 0.240, and 0.121, respectively. The dataset was created by BUPA Medical Research Ltd., and is publicly available at University of California at Irvine Machine Learning Repository ({\it https://archive.ics.uci.\\ edu/ml/datasets/Liver+Disorders}). 

\begin{figure}[!h]
\caption{Real application: boxplot of the empirical Youden index $J$ over 100 replications.}
\begin{center}
\includegraphics[width=0.75\textwidth]{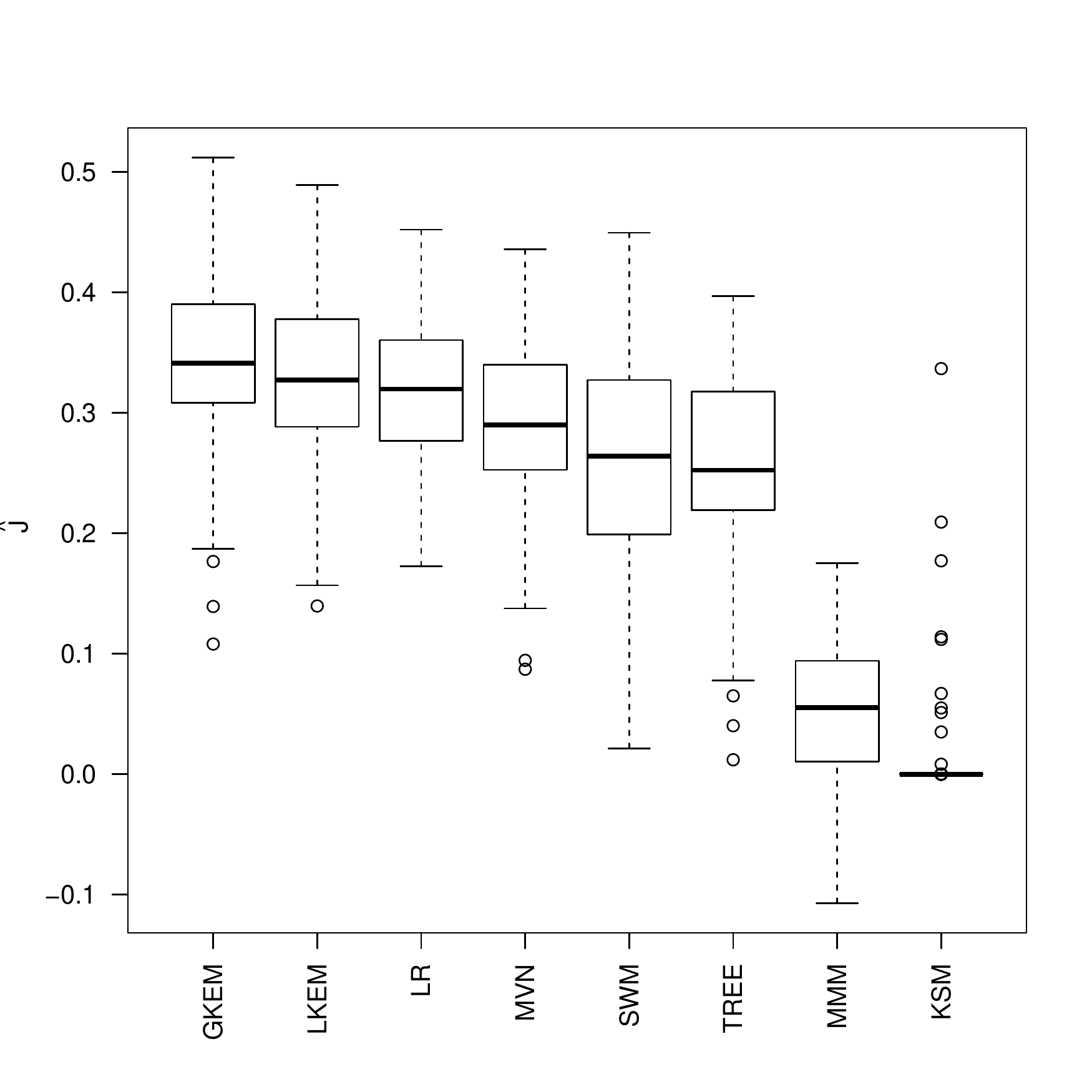}
\label{fig:Boxplot}
\end{center}
\end{figure}

The total 345 samples are randomly split into a training set of 200 samples and a testing set of 145 samples. We also set $\delta=0.1$  and select the tuning parameter $\lambda$ by 5-fold cross validation targeting on maximizing (\ref{eqn:empirical_J}). The experiment is replicated 100 times, and Figure \ref{fig:Boxplot} summarizes the averaged performance measures of our proposed method, MMM, MVN, KSM,  SWM, LR, and TREE.

\begin{center}
\begin{tabular}{c}
\hline
\hline
Figure \ref{fig:Boxplot} about here.\\
\hline
\hline
\end{tabular}
\end{center}

It is evident that our proposed method delivers competitive performance in comparison with other methods. It is also interesting to notice the significant improvement on diagnostic accuracy by combining biomakers nonlinearly. It is encouraging to note that our proposed methods with Gaussian kernel outperforms all other methods. 

\section{Closing remarks}

This paper proposes a flexible model-free framework for combining multiple biomarkers. As opposed to most existing methods focusing on the optimal linear combinations, the framework admits both linear and nonlinear combinations. The superior numerical performance of the proposed approach is demonstrated in a number of simulated examples and a real application to the liver disorder study, especially when the sample size is relatively large. Furthermore, the proposed method is especially efficient with a relatively large number of covariates present, where most existing methods relying on grid search are often inefficient. Further development could be on estimating confidence interval using perturbation resampling procedure \cite{Jiang2008} and combining biomarkers under covariate-adjusted Youden index setup \cite{Xu2014}.

\section*{Acknowledgment}
This research received no specific grant from any funding agency in the public, commercial, or not-for-profit sectors.

\section*{Appendix}

\noindent {\bf Proof of Proposition \ref{lem:consistency}.} Since $L_{\delta}(u)=L_{01} (u)+ \delta^{-1} (\delta-u)I(0 \leq u \leq \delta)$, we have
\begin{equation}
\begin{split}
E \Big ( w(Y) L_{\delta}( Y(g({\bf X})-c) ) \Big ) & =E \Big ( w(Y) L_{01}( Y(g({\bf X})-c)) \Big ) \\
& + E\Big(w(Y)\frac{\delta-Y(g({\bf X})-c)}{\delta} I(0 \leq Y(g({\bf X})-c) \leq \delta) \Big). \label{eqn:lemma2}
\end{split}
\end{equation}
Note that $E\big(w(Y) \frac{\delta-Y(g({\bf X})-c)}{\delta} I(0 \leq Y(g({\bf X})-c) \leq \delta) \big)$ is decreasing in $\delta$, and approaches 0 when $\delta \rightarrow 0$. Furthermore, for any given $\epsilon > 0$,
\begin{equation}
\begin{split}
E \big ( w(Y) L_{01}( Y(g(X)-c) ) \big ) - E \big ( w(Y) L_{01}( Y(g^*(X)-c^*)) & \\
= 
\int_{{\cal D}_{g,c,0} \cap {\cal D}^c_{g^*,c^*,0}} \frac{\pi-p(\bf{x})}{\pi(1-\pi)} f({\bf x}) d{\bf x} +
& \int_{{\cal D}^c_{g,c,0} \cap {\cal D}_{g^*,c^*,0}} \frac{p(\bf{x})-\pi}{\pi(1-\pi)} f({\bf x}) d{\bf x}\\
\geq 
\int_{{\cal D}_{g,c,\epsilon} \cap {\cal D}^c_{g^*,c^*,\epsilon}} \frac{\pi-p(\bf{x})}{\pi(1-\pi)} f({\bf x}) d{\bf x} +
& \int_{{\cal D}^c_{g,c,\epsilon} \cap {\cal D}_{g^*,c^*,\epsilon}} \frac{p(\bf{x})-\pi}{\pi(1-\pi)} f({\bf x}) d{\bf x}. \label{eqn:lemma2_2}
\end{split}
\end{equation}
By (\ref{eqn:bayes}), we have 
\[  \left\{\begin{array}{ll}
  \pi-p({\bf x}) > \epsilon, & \mbox{if ${\bf x} \in {\cal D}_{g,c,\epsilon} \cap {\cal D}^c_{g^*,c^*,\epsilon}$};\\
 p({\bf x}) - \pi > \epsilon, & \mbox{if ${\bf x} \in {\cal D}^c_{g,c,\epsilon} \cap {\cal D}_{g^*,c^*,\epsilon}$}.\\
\end{array}
\right.\]
Therefore, 
$$
E \big ( w(Y) L_{01}( Y(g^*_{\delta}(X)-c^*_{\delta}) ) \big ) - E \big ( w(Y) L_{01}( Y(g^*(X)-c^*)) > \epsilon Pr\left( {\cal D}_{g^*_{\delta}, c^*_{\delta},\epsilon} \triangle {\cal D}_{g^*,c^*,\epsilon}\right).
$$
By the fact that $E \Big ( w(Y) L_{\delta}( Y(g^*_\delta ({\bf X})-c^*_\delta) ) \Big ) \leq E \Big ( w(Y) L_{\delta}( Y(g^*({\bf X})-c^*) ) \Big )$, we have 
\begin{equation}
\begin{split}
E \big ( w(Y) L_{01}( Y(g^*_\delta(X)  & -c^*_\delta) ) \big ) - E \big ( w(Y) L_{01}( Y(g^*(X)-c^*))\\
\leq & E\Big(w(Y)\frac{\delta  -Y(g^*({\bf X})-c^*)}{\delta} I(0 \leq Y(g^*({\bf X})-c^*) \leq \delta) \Big). \\
\end{split}
\end{equation} 
It follows immediately that $Pr\left( {\cal D}_{g^*_{\delta}, c^*_{\delta},\epsilon} \triangle {\cal D}_{g^*,c^*,\epsilon}\right) \rightarrow 0$ as $\delta \rightarrow 0$. \hfill $\square$

%%%%%%%%%%%%%%%%%%%%%%%%%%%%%%%%%%%%%%%%%%%%%%%%%%%%%%%%%%%%%%%%%%%%%%%%

{}

\newpage

\begin{table}[!ht]
      \begin{center}
      \caption{Simulation examples: estimated means and standard deviations (in parentheses) of the empirical Youden index $J$ over 100 replications.}
      
      \medskip
        \begin{tabular}{rccc}
        \hline
        &~$n=100$~ & ~$n=250$~ & ~$n=500$~   \\
          \hline
            & \multicolumn{2}{c}{\it Example 1} & \\
           \hline
             LKME & 0.604 (0.0042) & 0.628 (0.0019) & 0.641 (0.0018) \\
             GKME & 0.572 (0.0063) & 0.604 (0.0029) & 0.623 (0.0023) \\
             MMM & 0.455 (0.0032) & 0.470 (0.0021) & 0.483 (0.0020) \\
             MVN & 0.633 (0.0018) & 0.638 (0.0014) & 0.647 (0.0012) \\
             KSM & 0.388 (0.0180) & 0.458 (0.0104) & 0.490 (0.0106) \\
             SWM & 0.555 (0.0065) & 0.594 (0.0044) & 0.611 (0.0035) \\
             LR & 0.628 (0.0022) & 0.639 (0.0017) & 0.646 (0.0017)  \\
             TREE & 0.490 (0.0068) & 0.525 (0.0047) & 0.559 (0.0029)  \\
               \hline
            & \multicolumn{2}{c}{\it Example 2} & \\
              \hline
			LKME & 0.636 (0.0075) & 0.690 (0.0025)  & 0.710 (0.0015) \\
            GKME & 0.612 (0.0054) & 0.654 (0.0045)  & 0.696 (0.0016)\\
            MMM & 0.609 (0.0033) & 0.622 (0.0025)  & 0.622 (0.0022)\\
            MVN & 0.573 (0.0065) & 0.571 (0.0047)  & 0.563 (0.0040) \\
            KSM & 0.214 (0.0281) & 0.046 (0.0164) & 0.047 (0.0171)\\
            SWM & 0.447 (0.0094) & 0.426 (0.0078) & 0.429 (0.0065)  \\
            LR & 0.648 (0.0054)  & 0.675 (0.0028)  & 0.678 (0.0025)  \\
            TREE & 0.433 (0.0052) & 0.512 (0.0039) & 0.555 (0.0036)  \\
           \hline
            & \multicolumn{2}{c}{\it Example 3} & \\
            \hline
            LKME & 0.296(0.0091)  & 0.367(0.0053)  & 0.389(0.0049)  \\
            GKME & 0.511(0.0052) & 0.568(0.0028)  & 0.592(0.0022)\\
            MMM &  0.423(0.0035) & 0.434(0.0021) & 0.443(0.0018)\\
            MVN & 0.344(0.0050)  & 0.371(0.0045)  & 0.377(0.0041) \\
            KSM & 0.192(0.0085)  & 0.193(0.0084)   & 0.202(0.0086)\\
            SWM & 0.370(0.0057) & 0.406(0.0028) & 0.417(0.0025)  \\
            LR & 0.307(0.0043) & 0.316(0.0030) & 0.320(0.0026) \\
            TREE & 0.424(0.0059)  & 0.477(0.0042) & 0.528(0.0031)  \\
            \hline
            & \multicolumn{2}{c}{\it Example 4} & \\
               \hline
               LKME & 0.103(0.0102)  & 0.150(0.0098)   & 0.209(0.0089) \\
               GKME & 0.529(0.0078) & 0.626(0.0050)  & 0.682(0.0028)\\
               MMM & 0.184(0.0084)  & 0.227(0.0034)   & 0.236(0.0026)\\
               MVN & 0.109(0.0071)  & 0.152(0.0056)  & 0.189(0.0054) \\
               KSM & 0.188(0.0050) & 0.213(0.0035) & 0.220(0.0028)\\
               SWM & 0.255(0.0078)  & 0.293(0.0050)   & 0.307(0.0039)  \\
               LR & 0.002(0.0023) & 0.004(0.0008)  & 0.011(0.0007)  \\
               TREE & 0.257(0.0143) & 0.364(0.0111)  & 0.368(0.0101) \\                                   
            \hline
        \hline
      \end{tabular}
       \label{tab:optimal_J}
       \end{center}
\end{table}

\end{document}